\documentclass[twocolumn]{iopart}
\usepackage{graphicx}
\begin{document}
\title{Damping of Rabi oscillations in quantum dots due to lattice
dynamics}
\author{Pawe{\l} Machnikowski and Lucjan Jacak}
\address{Institute of Physics, Wroc{\l}aw University of Technology,
50-370 Wroc{\l}aw, Poland}
\ead{Pawel.Machnikowski@pwr.wroc.pl}
%\maketitle

\begin{abstract}
We show that the interaction between carriers confined in a quantum dot
and the surrounding lattice under external driving of carrier dynamics
has a dynamical, resonant character. The quality of Rabi oscillations in
such a system depends on the relation between nonlinear spectral
characteristics of the driven dynamics and the spectral density of
effectively coupled lattice modes (phonon frequencies and density of
states). For a large number of Rabi oscillations within a fixed
time (allowed by e.g. exciton recombination) the spectrum of the dynamics
extends towards high frequencies, coming into resonance with acoustical
and optical phonons. Thus, this resonant
lattice response strongly restricts the possibility of fully coherent
control over the charge state in a quantum dot.
\end{abstract}

Demonstration of Rabi oscillations in semiconductor quantum dots (QDs)
recently reported by a few research groups
\cite{kamada01,stievater01,htoon02,zrenner02,borri02a} is a major
step towards coherent control of the charge state in these systems.
Such oscillations in a solid-state environment are strongly perturbed
by interaction with lattice modes.
It has been shown both experimentally
\cite{borri01} and theoretically \cite{krummheuer02,vagov02a,vagov03} that
coupling to phonons reduces the coherence of the charge system on
picosecond timescale. This effect may be related to lattice
relaxation following non-adiabatic (with respect to lattice modes)
optical excitation (phonon dressing effect) \cite{jacak03b}. It has
been shown \cite{alicki02a,alicki03} that this decoherence is a dynamical,
non-Markovian effect and may be minimized by slowing down the carrier
excitation dynamics.

In this paper we describe the dynamics of an exciton in an
InAs/GaAs QD
under coherent optical excitation in the presence of coupling to
lattice modes.
In order to correctly account for the dynamical properties of the
lattice, we solve the non-Markovian Master equation for the
reduced density matrix of the carrier subsystem (in Born
approximation, i.e. including one-phonon effects). The exciton states
in the QD are found numerically, including Coulomb interaction, in a
simple confinement model \cite{jacak03b} and neglecting
multi-exciton states. We find the dot occupation
(mean exciton number $\langle n\rangle$) upon exciting with a Gaussian
pulse of fixed duration and varying amplitude \cite{machnikowski03a},
parametrized by the rotation angle $\alpha$ on the Bloch sphere.
The proposed approach goes beyond the phenomenological description
used so far \cite{kamada01,borri02a} and yields insight into the
dynamical, resonant nature of the carrier--lattice interaction.

\begin{figure}[h]
\unitlength 1mm
\begin{center}
\begin{picture}(85,58)(0,5)
\put(0,0){\resizebox{85mm}{!}{\includegraphics{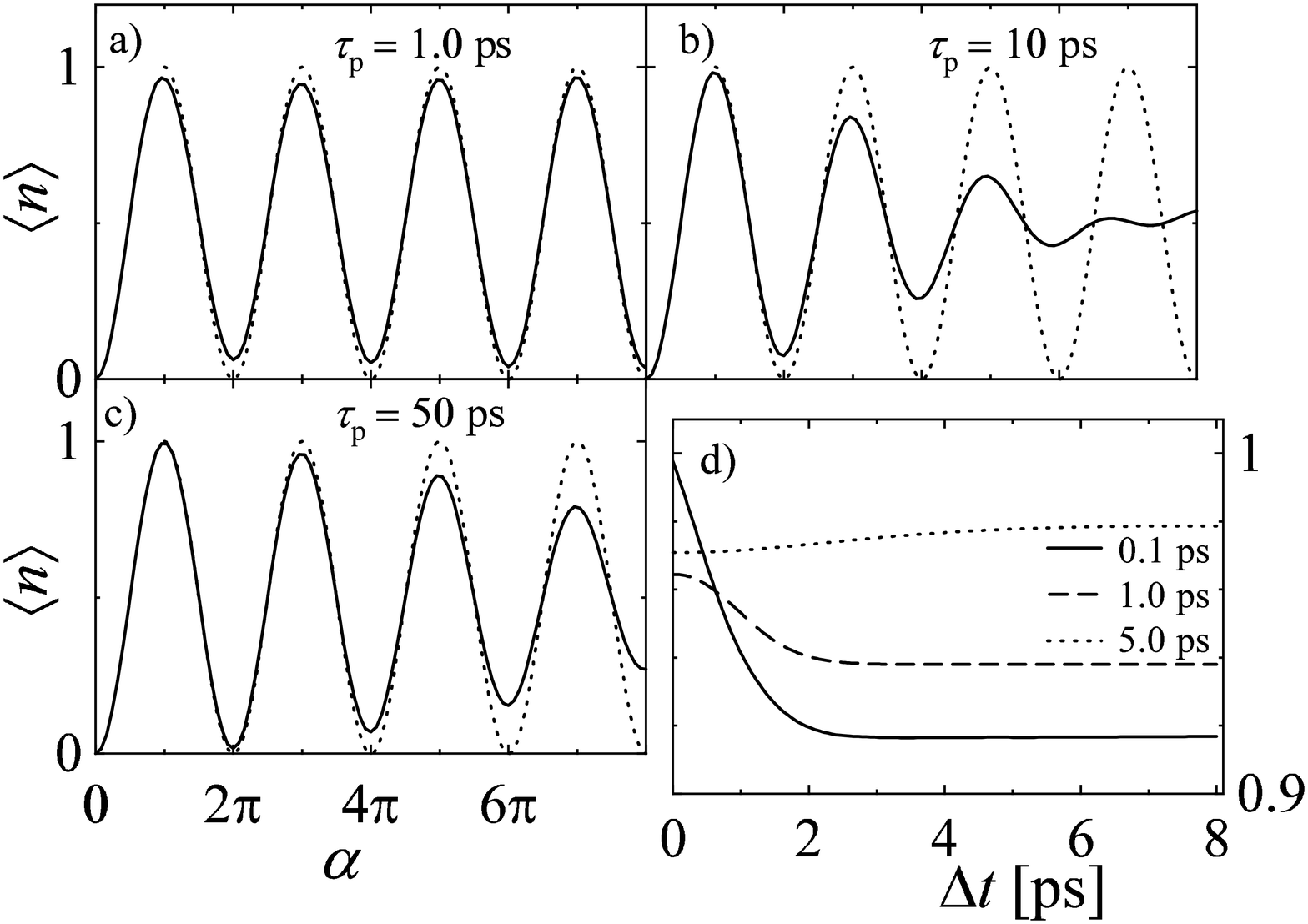}}}
\end{picture}
\end{center}
\caption{\label{rabi}(a-c) Pulse-intensity-dependent Rabi
oscillations (final QD occupation as a function of the rotation on
the Bloch sphere $\alpha$ performed by the pulse in the ideal
case) for various pulse durations as shown in the figure (solid)
compared to the ideal oscillation (dotted); (d) final QD
occupation after two $\alpha=\pi/2$ pulses separated by time
interval $\Delta t$ for pulse durations as shown; short pulses
create non-stationary (bare) states that loose coherence during
the delay, leading to a perturbation of the final state (cf.
\protect\cite{vagov02a}).}
\end{figure}

The resulting Rabi oscillations are shown on Fig. \ref{rabi}a-c (all
calculations are done for $T=10$ K).
Coupling to lattice modes reduces the amplitude of
these oscillations, the effect being small for short pulses
($\tau_{\mathrm{p}}=1$ ps) and increasing for longer pulses (10 ps), in
qualitative agreement with the experiment \cite{borri02a}. However,
for still longer pulses (50 ps), the quality of Rabi oscillation again
improves, manifesting the adiabatic character of the carrier-lattice
dynamics (transition through dressed states, as witnessed by their
lower decoherence in a two-pulse experiment, Fig. \ref{rabi}d).

An explanation of this behavior may be obtained by invoking the lattice
inertia, reflected by the natural timescales of phonon dynamics, i.e. by
phonon frequencies. If the induced carrier dynamics is much faster
than phonon oscillations the lattice has no time to react until the
optical excitation is done. The subsequent dynamics will lead to exciton
dressing, accompanied by emission of phonon packets, and will partly
destroy coherence of superposition states
\cite{borri01,krummheuer02,vagov02a,jacak03b} but cannot change
the exciton occupation number. In the opposite limit, the carrier dynamics
is slow enough for the lattice to follow adiabatically. The optical
excitation may then be stopped at any stage without any lattice relaxation
incurred, hence with no coherence loss (see Fig. \ref{rabi}d).
The intermediate case corresponds
to modifying the charge distribution in the QD with frequencies resonant
with the lattice modes which leads to increased interaction with phonons
and to decrease of the carrier coherence (cf. also \cite{axt99}).

\begin{figure}[h]
\unitlength 1mm
\begin{center}
\begin{picture}(85,58)(0,5)
\put(0,0){\resizebox{85mm}{!}{\includegraphics{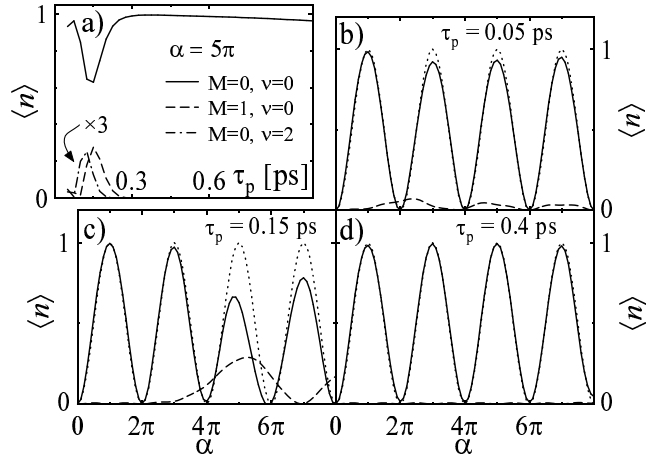}}}
\end{picture}
\end{center}
\caption{\label{opty}Final occupation of a few lowest exciton
states after an $\alpha=5\pi$ pulse as a function of the pulse
duration $\tau_{\mathrm{p}}$ (a) and pulse-intensity-dependent
Rabi oscillations for sub-picosecond pulse durations as shown
(b-d; solid: ground state occupation, dashed: 1st excited state
occupation, dotted: ideal oscillation). For
$\tau_{\mathrm{p}}=0.15$ ps, the frequency of induced dynamics
overlaps with the phonon frequencies, leading to large damping,
while for shorter (longer) durations it shifts towards higher
(lower) frequencies, decreasing the impact on the system dynamics.
The role of dark excited states ($M$ denotes angular momentum,
$\nu$ is another quantum number) is manifested by their non-zero
occupation.}
\end{figure}

More quantitatively, for growing number of rotations $m$,
the nonlinear spectral characteristics of the
optically induced dynamics develops a
series of maxima of increasing strength.
The position of the last and highest maximum corresponds approximately
to $2\pi m/\tau_{\mathrm{p}}$, in accordance with the resonance concept.
However, spectral components are also present at all the frequencies
$2\pi m'/\tau_{\mathrm{p}}$, $m'<m$, which is due to the turning on/off
of the pulse.
The interaction with lattice modes is strong when the frequencies of the
induced dynamics are in resonance with the natural lattice frequencies.
An interesting manifestation of this resonance effect may be observed
for sub-picosecond pulses, when the excitation of longitudinal optical (LO)
phonons becomes important (Fig. \ref{opty}).
Although the coupling between the ground state
and the LO modes is strongly reduced by charge cancellation, effects
involving higher (dark) exciton states may have considerable impact
\cite{jacak03b,fomin98}. The effect depends on the
relative positions of the narrow LO phonon features and of the induced
dynamics frequencies and has a non-monotonous character (Fig.
\ref{opty}a).

In conclusion, we have shown that the quality of Rabi oscillations of
exciton occupation in a quantum dot deteriorates due to resonant
excitation of lattice modes (overlapping nonlinear pulse spectrum and
phonon spectral density). Coherence may be protected from this effect by
slowing down the dynamics which, however, increases the impact of a number
of other decoherence mechanisms. Hence, the feasibility of coherent control
in such a system is strongly limited.

Supported by Polish KBN under Grants No. 2 P03B 02 424 and 2 P03B
085 25.

%\bibliographystyle{i:/init/tex/texmf-local/bibtex/bst/prsty}
%\bibliography{i:/init/tex/texmf-local/bibtex/bib/abbr,%
%i:/init/tex/texmf-local/bibtex/bib/quantum}

\end{document}